# Epitaxial Mn$_5$Ge$_3$ (100) layer on Ge (100) substrates obtained by flash lamp annealing

Yufang Xie[1,2*], Ye Yuan[1,3*], Mao Wang[1,2], Chi Xu[1,2], René Hübner[1], Jörg Grenzer[1], Yu-Jia Zeng[4], Manfred Helm[1,2], Shengqiang Zhou[1], and Slawomir Prucnal[1]

[1]*Institute of Ion Beam Physics and Materials Research, Helmholtz-Zentrum Dresden-Rossendorf, 01328 Dresden, Germany*

[2]*Technische Universität Dresden, 01062 Dresden, Germany*

[3]*Physical Science and Engineering Division, King Abdullah University of Science and Technology, 23955-6900 Thuwal, Saudi Arabia*

[4]*Shenzhen Key Laboratory of Laser Engineering, College of Optoelectronic Engineering, Shenzhen University, 518060 Shenzhen, P. R. China*

**Abstract:**

Mn$_5$Ge$_3$ thin films have been demonstrated as a promising spin-injector material for germanium-based spintronic devices. So far, Mn$_5$Ge$_3$ has been grown epitaxially only on Ge (111) substrates. In this letter we present the growth of epitaxial Mn$_5$Ge$_3$ films on Ge (100) substrates. The Mn$_5$Ge$_3$ film is synthetized via sub-second solid-state reaction between Mn and Ge upon flash lamp annealing for 20 ms at the ambient pressure. The single crystalline Mn$_5$Ge$_3$ is ferromagnetic with a Curie temperature of 283 K. Both the c-axis of hexagonal Mn$_5$Ge$_3$ and the magnetic easy axis are parallel to the Ge (100) surface. The millisecond-range flash epitaxy provides a new avenue for the fabrication of Ge-based spin-injectors fully compatible with CMOS technology.

* y.xie@hzdr.de, y.yuan@hzdr.de

In the past several decades, spintronic devices employing spin properties of electrons have been discussed as an intriguing alternative to conventional electronic charge devices.[1-5] The spin current in which carriers are highly spin-polarized strongly depends on the ferromagnetic properties of spin-injectors.[6-10] $Mn_5Ge_3$ films were epitaxially grown on Ge wafers with spin-polarization up to 42 %[11] and Curie temperature beyond room temperature enable fabrication of spintronic devices operated at room temperature.[12-14] Furthermore, the compatibility of transition metal germanides with group IV semiconductors enables the integration of spintronics with Si-based CMOS technolog.[13,15,16] Recently, it has been shown that a nanowire transistor made of $Mn_5Ge_3$/Ge/$Mn_5Ge_3$ exhibits a spin diffusion length of 480 ± 13 nm and a spin lifetime exceeding 244 ps.[14] To date, Mn-germanide is commonly observed in the form of Mn-rich Ge nanocrystals in the Ge matrix which has been expected to contain randomly distributed and dilute Mn impurities.[17-21] It was shown that the $Mn_5Ge_3$ films can be grown epitaxially only on Ge (111) wafers with out-of-plane magnetic easy axis.[22-27] The $Mn_5Ge_3$ on Ge (100) substrates grows in the form of either nano-islands[28-30] or polycrystalline microstructures which significantly limit the injection of spin-polarized current[31].

In the present work, we demonstrate the epitaxial growth of $Mn_5Ge_3$ (100) films on industry-relevant Ge (100) substrates. The $Mn_5Ge_3$ layer is approximately 40 nm thick with an extremely sharp interface between $Mn_5Ge_3$ and the Ge substrate. $Mn_5Ge_3$ (100) is obtained by diffusion of Mn into Ge during flash lamp annealing (FLA) performed for 20 ms. The fabricated epitaxial films exhibit ferromagnetism with a Curie temperature of $T_c$= 283 K. Moreover, in contrast to $Mn_5Ge_3$ on Ge (111), our sample shows an in-plane magnetic easy axis, which will contribute to the construction of spin-based devices.[15,32]

$Mn_5Ge_3$ films were grown by depositing a 30 nm thick Mn film on a Ge (100) wafer using electron-beam evaporation followed by FLA for 20 ms with an energy density of 95 $Jcm^{-2}$ at continuous $N_2$ flow. During a single flash pulse, Mn fully diffused into Ge forming a $Mn_5Ge_3$ epitaxial film. Structure analyses were performed by X-ray diffraction (XRD) and cross-sectional transmission electron microscopy (TEM). To obtain chemical information from the fabricated film, spectrum imaging analysis based on energy-dispersive X-ray spectroscopy (EDXS) was performed in scanning TEM (STEM) mode. The magnetic properties were studied by a superconducting quantum interface device equipped with vibrating sample magnetometer (SQUID-VSM). Magnetic-field-dependent Hall resistance was analyzed using van-der-Pauw geometry in a Lakeshore Hall measurement system in the temperature range of 2-300K.

Figure 1(a) displays a representative bright-field TEM image of the fabricated $Mn_5Ge_3$ film on the Ge substrate. It is evident that the investigated sample is composed of: (i) an ~25 nm thick polycrystalline top layer, (ii) an ~40 nm thick epitaxial interlayer and (iii) the single-crystalline Ge substrate. Fig. 1(b) shows a high-resolution TEM image obtained from the area marked by a white square in Fig. 1(a). Corresponding fast Fourier transforms (FFTs) calculated from the interlayer region as well as from the Ge substrate are depicted in Fig. 1(c) and (d), respectively. While the FFT from the face-centered cubic Ge substrate corresponds to a [1$\bar{1}$0] zone axis pattern, the FFT from the interlayer is well described based on hexagonal Mn$_5$Ge$_3$ in [010] zone axis geometry. Moreover, the FFT analysis clearly indicates the growth of Mn$_5$Ge$_3$ (100) with [001] in-plane axis parallel to the [110] axis of the Ge (001) substrate. Hence, the hexagonal c axis of Mn$_5$Ge$_3$ on Ge (100) is parallel to the film plane, while reported Mn$_5$Ge$_3$'s c axis on Ge (111) tends to be perpendicular to the film plane.[13, 23, 32] Importantly, the interface between the epitaxial Mn$_5$Ge$_3$ interlayer and the sample surface layer is rough, while the interface between Mn$_5$Ge$_3$ and the Ge substrate is very sharp (Fig. 1(a)). Note that neither remaining amorphous fractions nor second phase formation are observed at the interface.

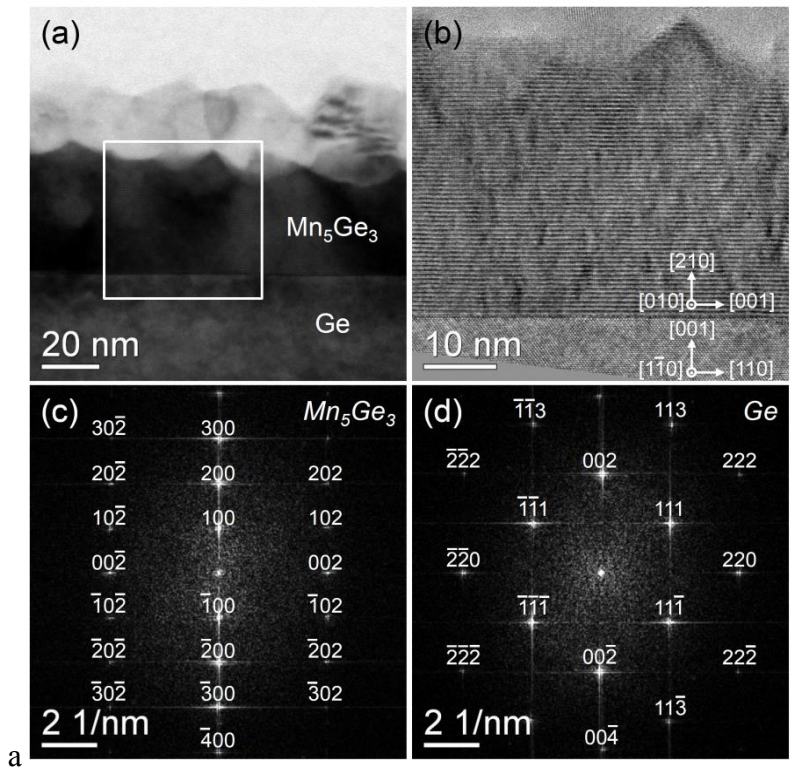

FIG. 1 (a) Cross-sectional bright-field TEM image of Mn$_5$Ge$_3$ (100) on Ge (100) substrate. (b) High-resolution TEM image of the area marked with a white square in (a). Fast Fourier transforms of hexagonal Mn$_5$Ge$_3$ in [010] zone axis geometry (c) and the cubic Ge substrate in [1$\bar{1}$0] zone axis geometry (d).

The room-temperature θ-2θ X-ray diffraction (XRD) pattern of Mn$_5$Ge$_3$ / Ge(100) film is shown in Figure 2. The prominent hkl reflections of Mn$_5$Ge$_3$ (100), (300) and (500) are observed together with the (200) and (400) reflections of the Ge substrate. The prominent peak positions of Mn$_5$Ge$_3$ coincide with the values of hexagonal Mn$_5$Ge$_3$ phase when the c-axis is oriented in the film plane.[33] The formation of hexagonal Mn$_5$Ge$_3$ also can be concluded from cross-sectional TEM analysis and the magnetic properties of the fabricated layer. According to Bragg's law, $2d\sin\theta = n\lambda$, where $\lambda$ is the wavelength of the incident X-rays ($\lambda_{CuK\alpha}$=1.54 Å), n is diffraction order and θ is the diffraction angle, the (100) lattice spacing is calculated to 6.21 Å which is related to the value (6.22 Å) of fully relaxed Mn$_5$Ge$_3$ (a=b=7.184; c=5.059 Å) in ICDD-PDF card 04-004-4927.[33] The peaks at 59.2° and 63.2° result from the diffraction at the Ge (400) plane by the X-ray excited from Cu K$_\beta$ edge and contamination irradiation from tungsten L$_\alpha$ edge, respectively.

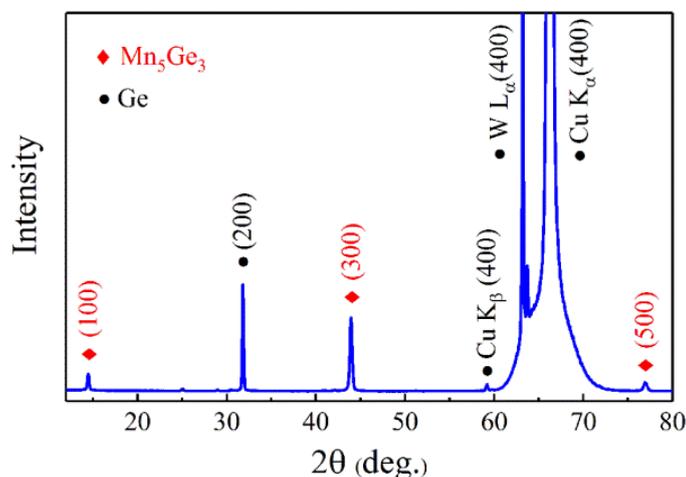

FIG. 2 XRD θ-2θ scan of epitaxial Mn$_5$Ge$_3$ (100) film grown on Ge (100) substrate. The diffraction peaks related to Mn$_5$Ge$_3$ are marked with solid diamonds and Ge substrate peaks are labeled with solid circles.

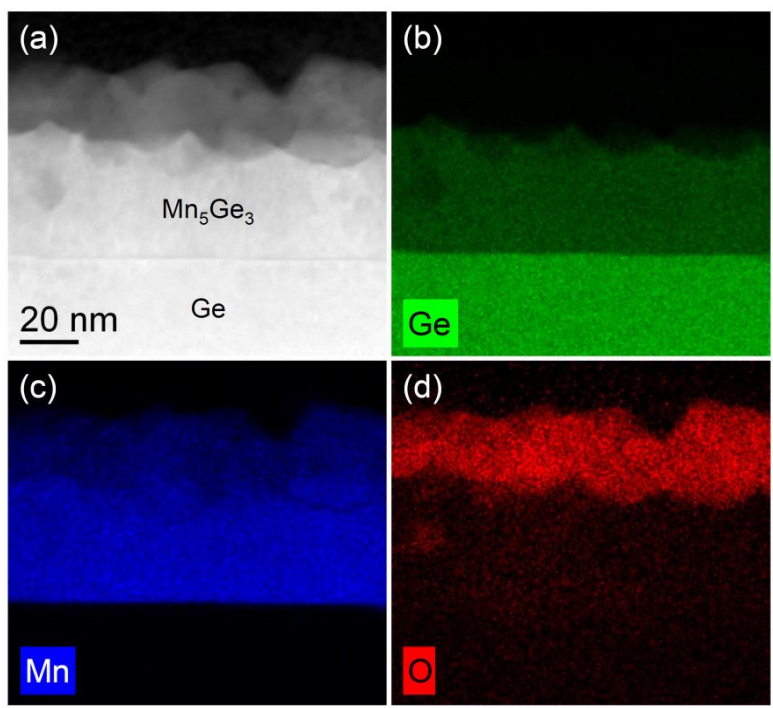

FIG. 3 (a) High-angle annular dark-field scanning TEM image of the Mn$_5$Ge$_3$ sample for the same specimen region as shown in Fig. 1(a) together with the corresponding element distributions of (b) germanium, (c) manganese, and (d) oxygen obtained by spectrum imaging based on energy-dispersive X-ray spectroscopy.

The chemical composition of the fabricated samples was investigated using spectrum imaging analysis based on energy-dispersive X-ray spectroscopy (EDXS) performed in scanning TEM (STEM) mode. Figure 3(a) presents a high-angle annular dark-field (HAADF)

STEM micrograph for the same region as shown in Fig. 1(a). Fig. 3(b), (c) and (d) show the element distributions of germanium - green, manganese - blue, and oxygen - red, respectively. The sample's top layer contains only Mn and O which indicates the formation of a manganese oxide. It is known that Mn films will be oxidized in the air without a protective layer.[34] In our case, the time interval between the Mn deposition and FLA is long enough for the formation of an about 25 nm thick Mn oxide. The oxidation of the Mn layer can be avoided by either integration of the FLA system within the metal deposition system and corresponding in-situ annealing after Mn film deposition or by adding a protective capping layer which can be selectively removed after FLA. According to Fig. 3, the interlayer consists of Mn and Ge with a composition of $Mn_{60}Ge_{40}$ which fully supports the above results about the formation of hexagonal $Mn_5Ge_3$. Interestingly, below $Mn_5Ge_3$, neither Mn nor O is found which, together with the high-resolution TEM analysis, suggests the formation of an extremely sharp interface between $Mn_5Ge_3$ layer and the Ge substrate.

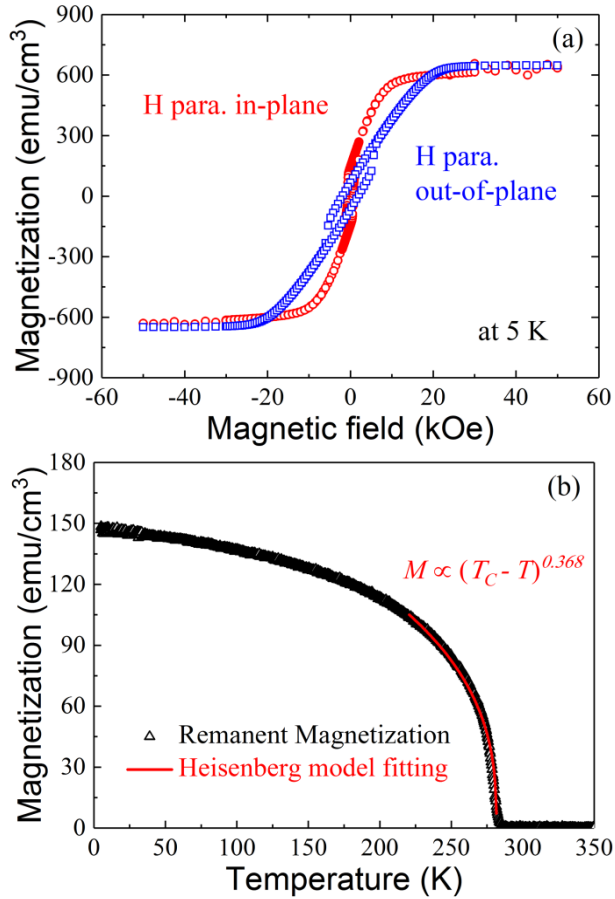

FIG. 4 (a) Magnetic field dependent magnetic hysteresis loops of $Mn_5Ge_3$ measured at 5 K when magnetic field is along the in-plane (circles) and the out-of-plane (squares) directions of the film plane, respectively. (b) Temperature dependent remnant magnetization under zero-field (black triangles). The red solid line shows a fitting curve by the Heisenberg theory.

Figure 4 shows the magnetic field and temperature dependent magnetization of the $Mn_5Ge_3$ sample. Fig. 4(a) shows the hysteresis loops obtained at 5 K after subtracting the diamagnetic background of Ge substrate. During the measurement, the magnetic field is along in-plane direction or the out-of-plane direction. A saturation field along the in-plane direction is much lower than along the out-of-plane direction (11 kOe vs. 23 kOe, respectively), implying that the in-plane direction is magnetic easy axis which is perpendicular to the hexagonal basal plane (parallel to c-axis). The $Mn_5Ge_3$/Ge (111) films with the same thickness shows out-of-plane magnetic easy axis, however still parallel to the hexagonal c-axis like our $Mn_5Ge_3$/Ge (100).[15, 32] The in-plane easy axis in $Mn_5Ge_3$/Ge (111) is observed only in very thin layers with d < 25 nm. The reorientation of the magnetic easy axis in $Mn_5Ge_3$/Ge (111) from in-plane to out-of plane is due to the shape anisotropy at the nanoscale.[15, 32] For a quantitative study of such magnetic anisotropy, the in-plane magnetic uniaxial anisotropy constant $K$ can be calculated as the energy difference per volume between the in-plane and the out-of-plane directions according to:

$$K = \frac{1}{V}\left(\int_{0[IP]}^{M_S} HdM - \int_{0[OOP]}^{M_S} HdM\right) \quad (1)$$

where $V$ is calculated volume, $M_S$ is saturation magnetization, $H$ and $M$ is the magnetic field and magnetization, respectively. The anisotropy constant is calculated as $2.8\times10^6$ erg/cm$^3$ at 5 K which is lower than the values of $3.5\times10^6$ erg/cm$^3$ for bulk sample at 77 K[35] and of $4.3\times10^6$ erg/cm$^3$ for $Mn_5Ge_3$/Ge(111) film at 15 K[32]. To determine the Curie temperature of fabricated film, the temperature dependent remnant magnetization was measured under a zero magnetic field during the warming process. After that the system was cooled down under a field of 2 kOe, and the result is displayed in Fig. 4(b). Upon gradually increasing temperature from 5 K, the magnetization decreases till 283 K where magnetization totally vanishes, indicating that the Curie temperature ($T_C$) is 283 K. It is worth noting that the $T_C$ herein is lower than the value (around 298 K) of $Mn_5Ge_3$ film on Ge (111) substrate,[36] probably due to the influence of the substrate orientation. Additionally, the magnetization presents a $(T_C - T)^\beta$ dependence with an exponent β of 0.368 slightly below $T_C$ (shown in the fitting curve in Fig. 4b in a red line), proving that the ferromagnetic magnetic behavior follows the short-range 3D Heisenberg theory at the magnetic critical regime approaching $T_C$.[37]

Figure 5(a) shows the result of the Hall resistance $R_{xy}$ as a function of magnetic field at 5 K when the field was perpendicular to the film plane. The Hall resistance $R_{xy}$ in magnetic materials is described by $R_{xy} = R_0H+R_SM$, where $R_0H$ is the ordinary Hall effect resulting from the Lorentz force and $R_SM$ is the anomalous term which is induced by magnetic contribution. Herein, the ordinary Hall component mainly appears beyond the saturation field (20 kOe) where the linear dependence with a positive slope indicates a *p*-type conducting behavior. However, the $R_{xy}$ curve shows a hysteresis behavior depending on the magnetic field, in particular at fields below 20 kOe, suggesting that the Hall resistance here is dominated by the anomalous term. In such regime, the Hall resistance is proportional to the magnetization $M$. The origin of anomalous Hall effect in magnetic materials can be classically described by $R_{AHE} = mR_{xx} + nR_{xx}^2$, where $R_{xx}$ is the longitudinal resistance and $m$ and $n$ are constants. As shown in Fig. 5(b), the Hall resistance $R_{xy}$ obeys a quadratic dependence as a function of $R_{xx}$, excluding the possibility that the origin of anomalous Hall effect in $Mn_5Ge_3$ is the skew scattering effect,[38] where the side jump scattering or intrinsic contribution dominates. However, these two contributions cannot be separated experimentally by dc measurements in theory.[38] Nevertheless, the anomalous Hall effect in $Mn_5Ge_3$ thin-film on Ge (111) substrate has been investigated by Zeng et al.,[36] and they claimed that the main contribution in $Mn_5Ge_3$ is intrinsic term.

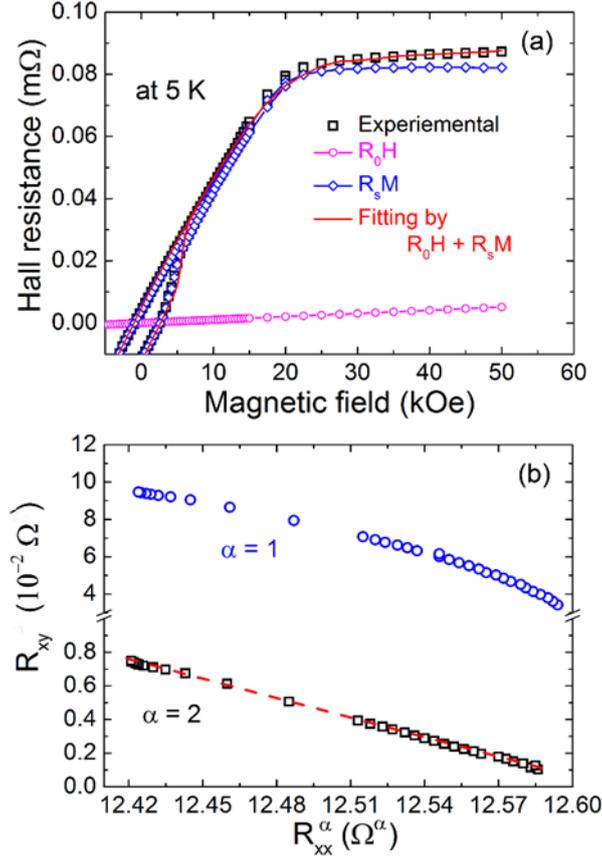

FIG. 5 (a) Experimental (black squares) and fitting (red solid line) results of magnetic field-dependent Hall resistance $R_{xy}$ which consists of ordinary (circles) and anomalous terms (diamonds). (b) The $R_{xy}$ as a function of $R_{xx}^\alpha$ for $\alpha = 1$ (circles) and $\alpha = 2$ (squares).

In conclusion, we report the epitaxial growth of a ferromagnetic $Mn_5Ge_3$ (100) layer on Ge (100) substrate by flash lamp annealing. An extremely sharp interface between $Mn_5Ge_3$ and the Ge substrate is observed, ensuring possible high spin transfer efficiency. The magnetic hysteresis loops, 3D-Heisenberg-like remnant magnetization curve as well as anomalous Hall Effect confirm the formation of a ferromagnetic layer with a Curie temperature of 283 K. Moreover, the magnetic easy axis lies in the film plane, which is promising for realizing high-efficient spin injection. Our work lightens an avenue of applying $Mn_5Ge_3$ layers as spin-injector, particularly when it is combined with industrial Ge (100) wafer-based technology.

**Acknowledgement**

This work is financially supported by the Helmholtz Association of German Research Centers (HGF-VH-NG-713). The author Y. Xie (File No. 201706340054) thanks the financial support by China Scholarship Council. Support by A. Scholz is gratefully acknowledged. Furthermore, the use of HZDR Ion Beam Center TEM facilities and the funding of TEM Talos by the German Federal Ministry of Education of Research (BMBF), Grant No. 03SF0451 in the framework of HEMCP are acknowledged.